\documentclass[11pt,twoside]{article}
\pdfoutput=1
\usepackage{./asp2014}

\aspSuppressVolSlug
\resetcounters

\begin{document}

\title{IVOA Data Access Layer: Goals, Achievements and Current Trends}
\author{Fran\c cois~Bonnarel,$^1$ Patrick~Dowler,$^2$  Keith~Noddle$^3$and Douglas Tody$^4$}
\affil{$^1$CDS, Strasbourg, France; \email{francois.bonnarel@astro.unistra.fr}}
\affil{$^2$CADC, Victoria, British Columbia, Canada; \email{patrick.dowler@nrc-cnrc.gc.ca}}
\affil{$^3$IfA, University of Edinburgh, Edinburgh, UK; \email{keith@keithnoddle.org}}
\affil{$^4$NRAO, Socorro, New Mexico, USA; \email{dtody@nrao.edu}}
\paperauthor{Sample~Author1}{Author1Email@email.edu}{ORCID_Or_Blank}{Author1 Institution}{Author1 Department}{City}{State/Province}{Postal Code}{Country}
\paperauthor{Sample~Author2}{Author2Email@email.edu}{ORCID_Or_Blank}{Author2 Institution}{Author2 Department}{City}{State/Province}{Postal Code}{Country}
\paperauthor{Sample~Author3}{Author3Email@email.edu}{ORCID_Or_Blank}{Author3 Institution}{Author3 Department}{City}{State/Province}{Postal Code}{Country}
%

\section{Introduction}
The Virtual Observatory can be defined as a set of standard protocols useful for interoperability of data, data archives, and applications. These standards facilitate the definition of shared datamodels for astronomical data of all kinds, shared transport formats and standard registration of on-line astronomical resources. Standard interfaces are also defined for applications and services. Data Access Layer protocols are an important part of the whole scheme. They standardize the way applications can query data services, enable data discovery, describe resources and define data access and retrieval methods via server side processing. In 2002 the IVOA created a dedicated working group (the Data Access Layer Working Group or DAL) to define and develop these universally accepted standards.

     DAL protocols provide solutions to questions such as :
"\textit{Describe in a standard way all datacubes containing the 21 cm line in an area of 1 deg radius around quasar 3C 273}"."\textit{Retrieve the central part of one of this cube giving the spatial and spectral size of the sub-cube designated by a standard identifier}"."\textit{Retrieve from all catalog services in the VO all the sources of all stellar objects around the center of Crab nebula}"'. 

The core work of the DAL WG is the definition of interface specifications. DAL defines standards allowing the development of interoperable services but without dictating implementation details. The DAL protocols define the way to query services and the response to be returned.

DAL protocols have strong relationships with other IVOA standards. \textbf{Query} operations and responses are based upon IVOA \textbf{DataModel} concepts. Usage of \textbf{DAL standardized vocabulary} is consistent with work done by the IVOA \textbf{semantics} working group. Standardized \textbf{description} and \textbf{discovery} of the \textbf{services} is done according to rules defined by the IVOA \textbf{Registry} protocols. \textbf{Asynchronous} mode support of the DAL services rely on functionalities defined by the IVOA \textbf{Universal Worker Service} (UWS). Eventually \textbf{VO Applications} are consumers of DAL services.

\section{First generation of DAL specifications: ConeSearch, SIA1.0 }
One of the first successes of IVOA was the definition and recommendation of a TABLE transport protocol, \textbf{VOTable} based on a specific XML schema (\cite{2011arXiv1110.0524O}). Early DAL protocols were HTTP GET services with predefined input parameters and a specific output TABLE.
\textbf{ConeSearch} ( \cite{2011arXiv1110.0498W} ) standardized "region of interest" (ROI) constraints for catalog of sources, whilst
\textbf{SIA1.0} (\cite{2011arXiv1110.0499T} ) was devoted to discovery and retrieval of images and cubes. The Query phase used ROI constraints like ConeSearch, plus additional constraints. The VOTable response describes the images and cubes in a standard way. An implicit image data model (based upon the FITS standard) underlies this image description. Virtual data generation (cutouts and mosaics) are also possible as well as image retrieval. 
While quite successful and simple to implement and use (11785 ConeSearch services and 195 SIA1.0 services are available from the IVOA registry as of October 2015) this first generation of standards predates modern IVOA standards and is in the process of being replaced by newer protocols based upon updated service semantics and datamodels.

\section{2nd generation protocols : TAP, ADQL, SSA and SLA} 
The Astronomical community makes large use  of measurements stored in catalogs. Catalogs are searched, exchanged, cross-identified and plotted in various ways to make science. 
 The relational model is well suited to describing attributes and relationships between the tables. Relational DBMS implement this model in various ways. The goal of the IVOA is to offer an interoperable common distant interface to services of tables and catalogs implemented in various relational DBMS. A first and unsuccessful attempt was made with \textbf{Skynode} using the \textbf{VOQL} query language. The \textbf{TAP protocol} (\cite{2011arXiv1110.0497D}) offers this functionality by exposing the database schema in a standardized way with TAP\_SCHEMA and VOSI-tables description. It exposes the metadata of the tables provided by the service (table and column names, units, datatypes and extended types, IVOA Uniform Content Descriptor and Universal Types).  Queries can be synchronous or asynchronous. Asynchronous queries rely on the IVOA \textbf{UWS }architecture ( \cite{2011arXiv1110.0510H}). TAP queries can be expressed in different languages in principle, but most of the TAP services exclusively use the mandatory \textbf{Astronomical Data Query Language} (ADQL - \cite{2011arXiv1110.0503O}) which is an extension of SQL enriched by astronomy specific features  for client requests. There are 35 TAP services in the IVOA registry at the current date of October 2015. 

Spectra and "Spectral Energy Distribution" are specific kind of datasets very common in Astronomy. They occur each time the spatially integrated light emission of astronomical sources is spectrally analyzed and differentiated. 
 The IVOA is driven by scientific imperatives to deliver standardized ways to exchange spectra. The Data model Working group delivered the \textbf{Spectrum Data Model} (\cite{SPECTRUM}), an abstract and exhaustive description of measurements and metadata within a spectrum which can be serialized in several ways. The Spectrum Data Model reuses some parts of the \textbf{STC Data Model} (\cite{2011arXiv1110.0504R} ) and of the\textbf{ Characterization Data Model} (\cite{2011arXiv1111.2281L}). \textbf{Simple Spectral Access protocol} (SSA - \cite{SSA}) uses standard query parameters to constrain all axes of the data as well as more specific features of the spectrum model. The response is an organized VOTable where GROUPS are mapped from the Spectrum data model packages and where the Table elements  are tagged as model attributes using the utype attribute. SSA protocol includes virtual data generation by server-side processing based on a "best match to query parameters" delivery strategy.

The \textbf{Simple Line Access} (SLA - \cite{2011arXiv1110.0500S} ) protocol allows retrieval of spectral lines as VOTABLE serializations   of the \textbf{Simple Spectral Line Data Model}. This standard follows the same principles as SSA.

\section{Recent or current developments of new protocols driven by new science needs} 
Recently the avalanche of terabytes of new astronomical data produced by  current or planned survey projects in all wavelengths reinforced the need for developing protocols to discover, describe and access all kind of datasets more efficiently. 

    A TAP technology based upon a generic dataset observation-summary metadata model called \textbf{ObsCore} lead to the emergence of \textbf{ObsTAP} services (\cite{2011arXiv1111.1758L} ), allowing the discovery of observational datasets of all kinds: raw or calibrated images, cubes, spectra, event lists, etc...

     In parallel the IVOA Committee of Science Priorities encouraged the Working group to provide more specific protocols adapted to multidimensional data as new large observational projects urgently need to provide open access and interoperability for cross-correlation of their data. In this context, \textbf{DataLink} (\cite{2015arXiv150906152D} ) defines various methods and technologies to link on-line additional resources useful for better usage to known or already discovered datasets. \textbf{SIAV2.0} is a parameterised query (PQL) adapted to multidimensional data of the ObsCore 1.1 data model for simple data discovery. \textbf{AccessData} is driving server-side processing for extracting information from the datasets. Intended applications are cutout, filtering, re-sampling or re-gridding and combination of several datasets. IVOA DAL manages a large number of protocols implemented in many services. Many of the questions faced in developing these protocols are common questions, such as Query global syntax, management of valid and error queries, format of the responses, management of synchronous or asynchronous modes, availability or capability auto-description. The common patterns and features have been extracted into \textbf{Data Access Layer interface} (DALI - \cite{2014arXiv1402.4750D} ) specification for easy re-use and consistency.

\section{Conclusion: Lessons learned, open questions}

   The use cases for accessing astronmical data are so disparate that it is infeasible to provide one single service protocol satisfying all the requirements and allowing all capabilities, hence the approach has been to define distinct but consistent protocols for the different classes of data.
   A complete solution will only be achieved when all current protocols are realised. In this context DALI helps to federate development of such combined services.
   However two branches of protocols have existed for several years and will still coexist and "collaborate" in the future.
   Relational-model protocols (TAP, ADQL, ObsTAP) allow the most complex queries and are well suited for generic discovery of astronomical datasets. However, they are complex to implement for data archives. At the same time object-model protocols historically called "simple" protocols (Cone Search, SIA version 1 and 2, SSA, SLA) are simpler and more flexible to implement and can be more user friendly. However they require a solid and well defined underlying data model for each class of astronomical data.
   The discovery and data access (e.g., AccessData) protocols currently being implemented will be integrated to provide capabilities ranging from discovery to advanced data access via virtual data generation. At present DataLink already allows a flexible set of combinations between protocols of the two branches (eg, AccessData functionalities can be attached to either ObsTAP services or to "Simple" access services using any of these techniques).
    The Parameter Query Language evolution remains an open question because it is difficult to balance user-friendly readability of the syntax with the homogeneity of the input query content and the VOTable query response. 
     At the response level, utypes have proven to be very efficient to link various level elements in VOTable responses with VO DataModels attributes and class definitions and are plainly efficient for metadata description and discovery tasks. They are less efficient for full VO serialisation of datasets in VOTable, which is the reason why the DataModel working group proposed the \textbf{"VO-DML mapping to VOTable"} for full inclusion of the DM structures into the responses. Again it's still an open question to know how these innovations will benefit future versions of the DAL protocols.

\acknowledgements The authors are the four successive DAL chairs in reverse chronological order. They want to thank specifically the four successive DAL working
group vice-chairs: M.Molinaro, M.Fitzpatrick, J.Salgado and M.Dolensky and further
more all the colleagues participating to the DAL WG as well as members
of other IVOA WG. F.Bonnarel acknowledges support from the Astronomy
ESFRI and Research Infrastructure Cluster - ASTERICS project, funded by
the European Commission under the Horizon 2020 Programme (GA 653477). 

\bibliography{p012}

\end{document}